\documentclass[reprint,aps,prb,superscriptaddress,showpacs,longbibliography]{revtex4-1}

\usepackage[pdftex]{graphicx} 
\usepackage{float}
\usepackage{amsmath} 
\usepackage{amsfonts} 
\usepackage{amssymb}
\usepackage[pdftex,colorlinks=true]{hyperref}
\usepackage{mathtools}

\DeclarePairedDelimiterX\expval[1]{\langle}{\rangle}{#1}

\usepackage[colorinlistoftodos,prependcaption,textsize=tiny]{todonotes}

\def\bea{\begin{eqnarray}}
\def\eea{\end{eqnarray}}
\def\beq{\begin{equation}}
\def\eeq{\end{equation}}

\AtBeginDocument{%
    \newwrite\bibnotes
    \def\bibnotesext{Notes.bib}
    \immediate\openout\bibnotes=\jobname\bibnotesext
    \immediate\write\bibnotes{@CONTROL{REVTEX41Control}}
    \immediate\write\bibnotes{@CONTROL{%
    apsrev41Control,author="08",editor="1",pages="1",title="0",year="1"}}
     \if@filesw
     \immediate\write\@auxout{\string\citation{apsrev41Control}}%
    \fi
}%

\begin{document}

\title{Rod motifs in neutron scattering in spin ice}

\author{Claudio Castelnovo} \affiliation{TCM Group, Cavendish Laboratory, 
University of Cambridge, J.~J.~Thomson Avenue, Cambridge CB3 0HE, 
United Kingdom}

\author{Roderich Moessner} \affiliation{Max-Planck-Institut f\"{u}r Physik
komplexer Systeme, N\"othnitzer Str.\ 38, 01187 Dresden, Germany}

\begin{abstract}
In classical and quantum spin ice, rod-like features appear in the neutron 
scattering structure factor when the pinchpoints characteristic of classical 
spin ice get washed out. 
We show that these features do {\it not} indicate the absence 
of spin correlations between planes perpendicular to the rods. 
Rather, they arise because neutron scattering is largely 
insensitive to the three-dimensional correlations which are present 
throughout. We present two very simple models 
which exhibit a pristine incarnation of such scattering rods. 
This provides a physical picture for their appearance, 
elucidates the role played by monopole excitations and identifies 
conditions conducive to their observation. 
\end{abstract}

\maketitle
%
%

Spin ice~\cite{BramwellGingrasSciencereview} 
is arguably the only three dimensional magnetic material for
which fractionalisation and the presence of an emergent gauge field
have been established~\cite{Castelnovo2012}. 
Neutron scattering studies have played a
central role, not least in pinning down qualitatively important
features such as the pinch 
points~\cite{IsakovDipolarCorrelations,Fennell2009,Kadowaki2009} 
and Dirac strings~\cite{CMSMonopoles,Morris2009}, 
indicating the presence of an emergent gauge structure. 

Recently, many studies of potential quantum spin ice compounds have
extended these investigations. One resulting observation is that in
these systems, the pinchpoints are less well-defined. This is an
important observation, as it is believed to be related to the
presence of fractionalised monopole excitations: these are sources and
sinks of the emergent gauge field, and as such degrade the
conservation law responsible for the appearance of the pinch points.
\begin{figure*}[ht!]
\includegraphics[height=0.25\textwidth]{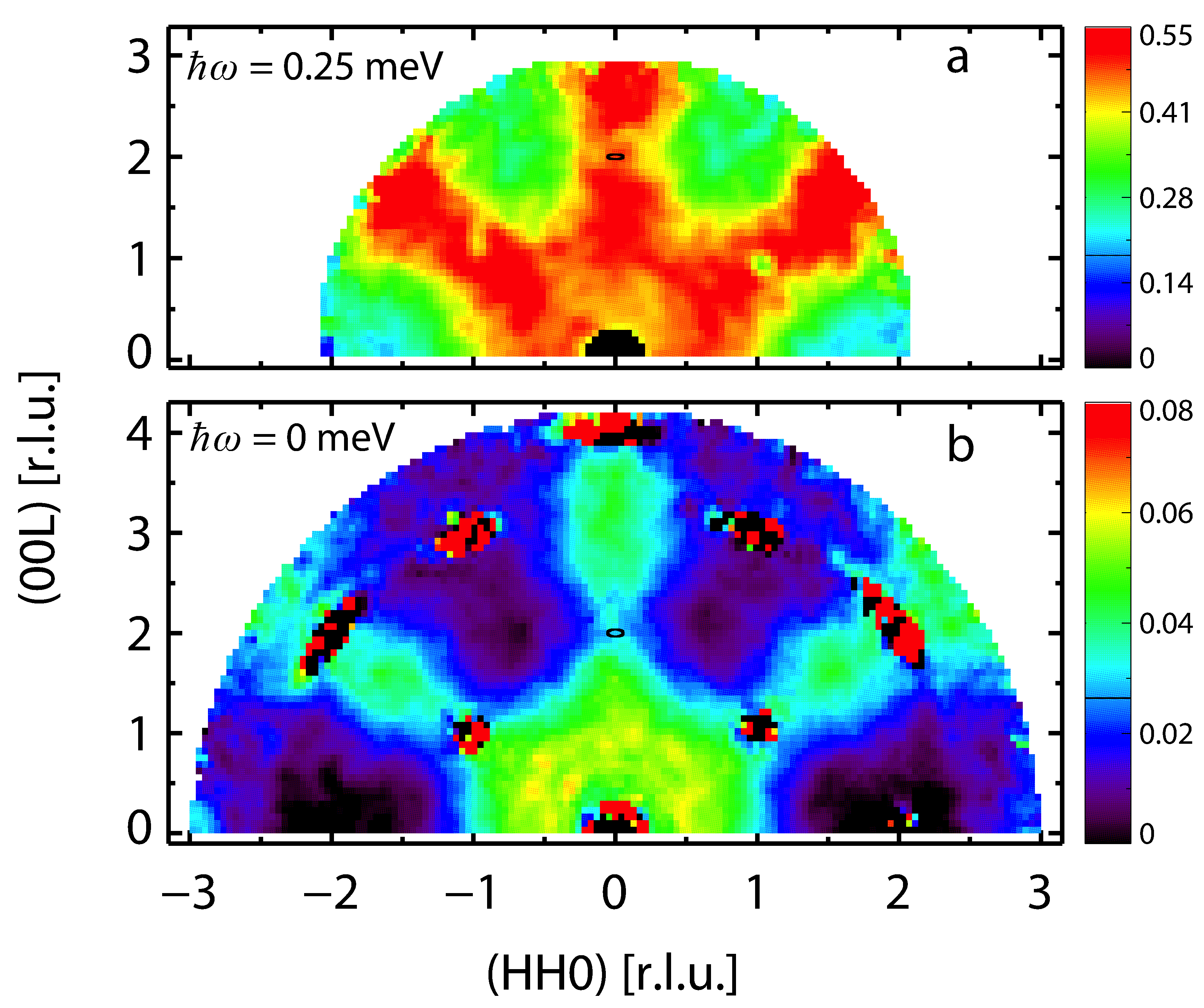}
\hspace{0.1 cm}
\includegraphics[height=0.25\textwidth]{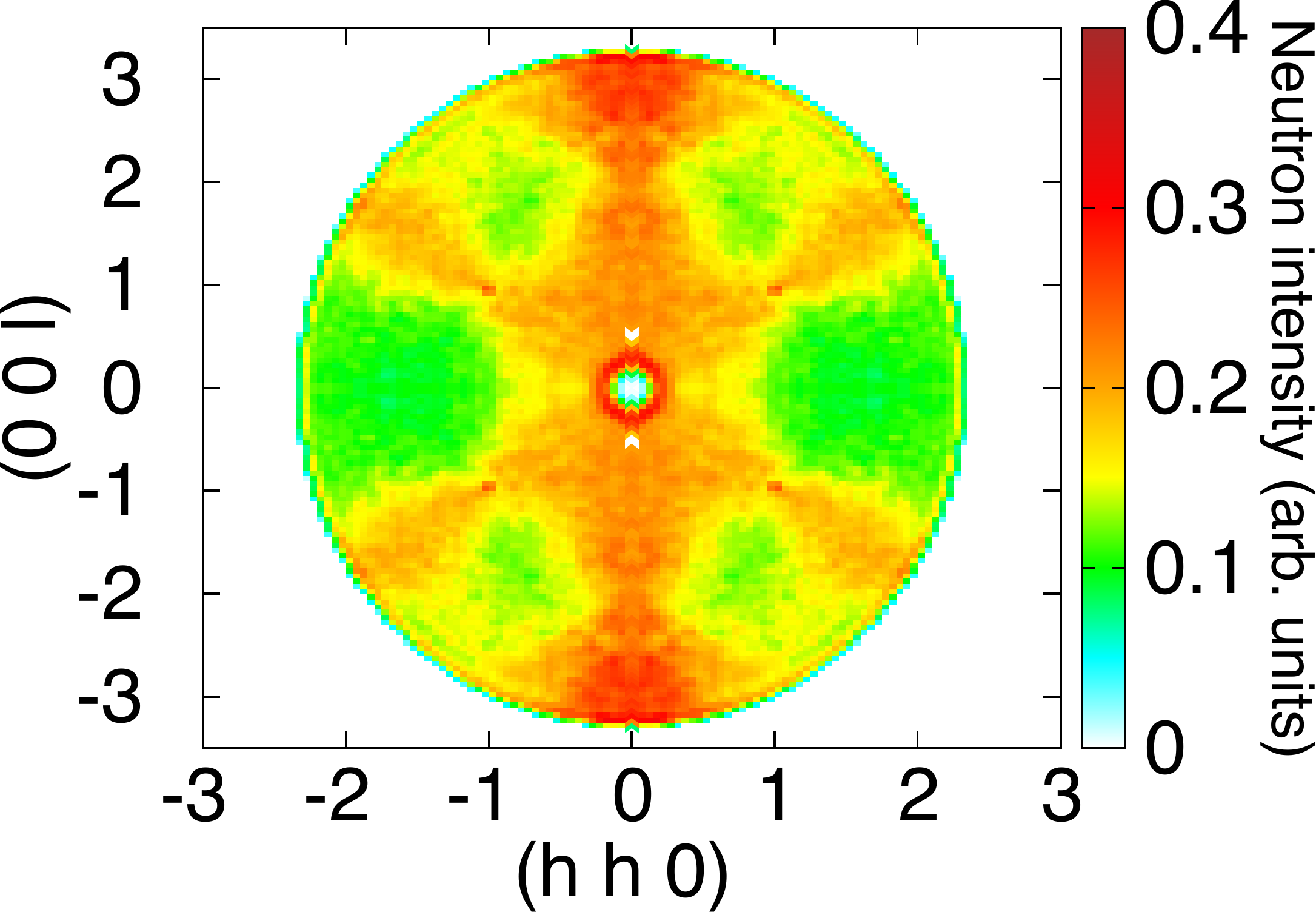}
\hspace{0.2 cm}
\includegraphics[height=0.25\textwidth]{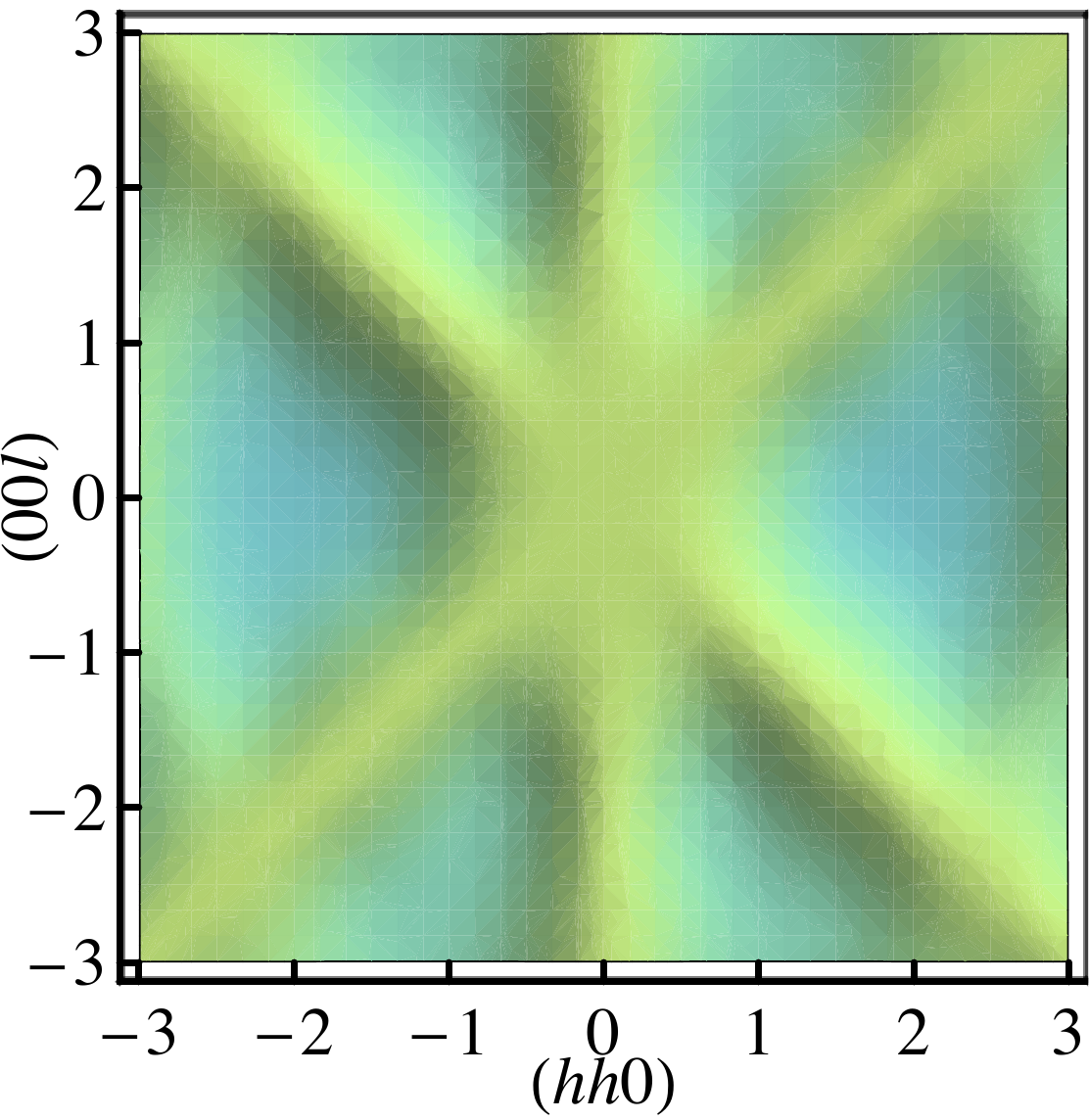}
\caption{
\label{fig: rods literature} 
Rod features in neutron scattering in the [200] and [111] 
directions in reciprocal space [where (r.l.u.) represents reciprocal lattice units]. 
Left panel: inelastic and elastic neutron scattering (top/bottom portion of the panel) 
on Pr$_2$Zr$_2$O$_7$ from Ref.~\onlinecite{Kimura2013}. 
Middle panel: inelastic neutron scattering on Nd$_2$Zr$_2$O$_7$ from 
Ref~\onlinecite{Petit2016}. 
Right panel: neutron scattering structure factor obtained from quantum 
Monte Carlo simulations of quantum spin ice (XXZ Hamiltonian) from 
Ref.~\onlinecite{Kato2015}. 
}
\end{figure*} 

As pinch points get washed out, they  leave
behind somewhat uniform, one-dimensional motifs, 
as observed in the classical spin ice compound
Dy$_2$Ti$_2$O$_7$~\cite{Fennell2009}, as well as the 
candidate quantum spin ices Pr$_2$Zr$_2$O$_7$~\cite{Kimura2013}
and Nd$_2$Zr$_2$O$_7$~\cite{Petit2016,Petit2016a}, with concomitant
observations in quantum Monte Carlo simulations~\cite{Kato2015};
representative plots are reproduced in Fig.~\ref{fig: rods literature} from 
Refs~\onlinecite{Kimura2013,Kato2015,Petit2016}. 
Such scattering rods are not unheard of in pyrochlore
compounds -- indeed, for Gd$_2$Ti$_2$O$_7$ rods have also been sighted
and successfully related to the presence of uncorrelated planes, whose
normal is given by the rod direction~\cite{Raju1999}.

This raises the twin questions of (i) how to account 
for the role of monopoles in the genesis of the scattering
rods and (ii) how their presence leads to decorrelation between
planes in the spin ice context.

The answer to the second question is that the rods do not in fact indicate 
decorrelated planes. Rather, three-dimensional 
correlations persist which are hidden by the matrix elements
characteristic of the neutron scattering process. We demonstrate this
by presenting two very simple models, neither of 
which exhibits decoupled planes but which do show rods in neutron 
scattering: the first is studied in Monte Carlo simulations while the second 
is a fully tractable analytical calculation for a single tetrahedron.
The numerical model has the absence of doubly-charged monopoles as its sole 
ingredient, while tetrahedra without, or with a singly-charged, monopole are 
treated as degenerate. The single-tetrahedron analytics 
straightforwardly relates the rods to properties of the classical 
interaction matrix for spin ice. 

This demonstrates that the rod-like features persist in a regime far 
removed from the dilute monopole limit, where an emergent gauge field is a  
natural degree of freedom; their role is above all to 
decorrelate spins in neighbouring tetrahedra from one another, so that 
the single tetrahedron result agrees with that of the full lattice with 
doubly-charged monopoles suppressed. 

We discuss the settings in which this condition is approximately met. 
These include quantum spin ice materials at intermediate energies, as well
as classical spin ice at intermediate temperatures. 
%
%

{\it Models and results.} 
We first show results from Monte Carlo simulations of classical 
dipolar and nearest-neighbour spin ice, 
\beq
H = - J \sum_{\langle i,j \rangle} \vec{S}_{i}\cdot\vec{S}_j 
+ D \sum_{i,j} 
\left[ 
  \frac{\vec{S}_{i}\cdot\vec{S}_j}{\vert \vec{r}_{ij} \vert^3} 
	- 
	\frac{3 (\vec{S}_{i}\cdot\vec{r}_{ij})(\vec{S}_j\cdot\vec{r}_{ij})}
	     {\vert \vec{r}_{ij} \vert^5}
\right] 
\eeq
as well as single tetrahedron calculations. 
For the nearest-neighbour case we choose $J=9$~K and $D=0$, whereas for 
the dipolar case we use $J=-3.72$~K and $D=1.41$~K; the distances 
$\vec{r}_{ij}$ are measured in units of the pyrochlore nearest-neighbour 
distance. 

\begin{figure*}[ht!]
\includegraphics[height=0.182\textwidth]{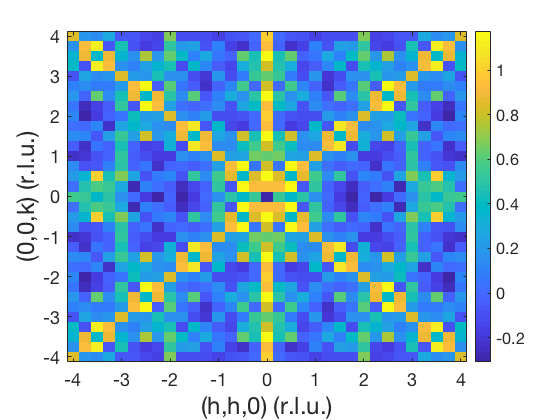}
\includegraphics[height=0.182\textwidth]{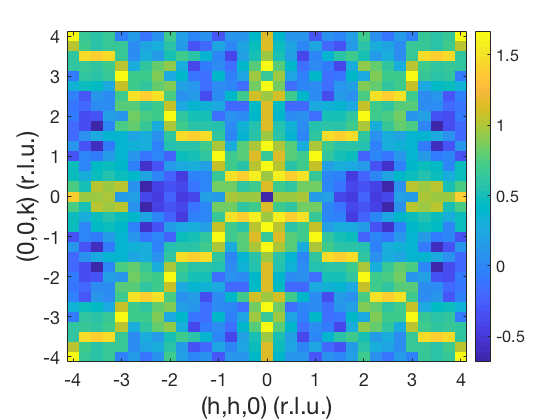}
\includegraphics[height=0.182\textwidth]{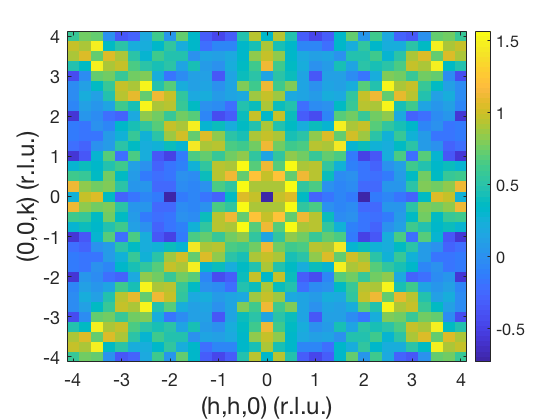}
\includegraphics[height=0.182\textwidth]{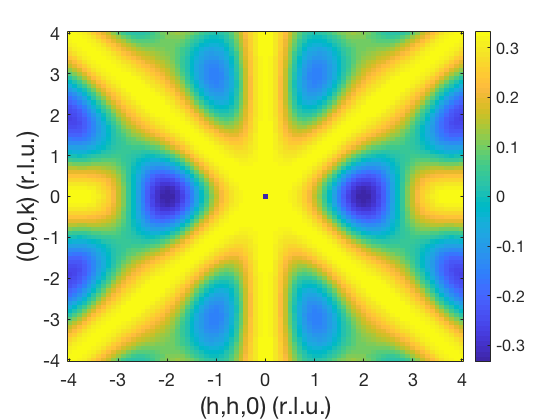}
\caption{
\label{fig: n2m} 
Rods in neutron scattering resulting from four different models:
First/second panel: Monte Carlo simulations for $L=8$ on dipolar 
($J=-3.72$~K, $D=1.41$~K, $T=10$~K) and
nearest-neighbour ($J=9$~K, $T=10$~K) spin ice;
Third panel: Stochastic sampling over spin ice configurations where 
double monopole tetrahedra are strictly forbidden, also for $L=8$. 
Right Panel: Single tetrahedron calculation, summed uniformly over all 
configurations but the all-in and all-out ones. 
}
\end{figure*}

We compare: 
(i) Monte Carlo simulations for the nearest-neighbour and dipolar spin ice 
models at high temperature; 
(ii) simulations of non-interacting spins subject to the hard constraint 
that all-in and all-out tetrahedra are strictly forbidden 
(no double monopoles); and 
(iii) a single tetrahedron calculation that accounts only for 2in-2out and 
single monopole configurations (i.e., double monopoles are forbidden). 
The results, in the form of neutron scattering structure factor, 
\beq
\mathcal{F} = \!\!\!\!\!\!
\sum_{\alpha,\beta = x,y,z} 
\left(
  \delta_{\alpha\beta}
	- 
	\frac{k_\alpha\, k_\beta}{k^2}
\right)
\sum_{i,j} e^{\i \vec{k}\cdot\vec{r}_{ij}} 
  \langle S^\alpha_i S^\beta_j \rangle 
\, , 
\label{eq: NSF full}
\eeq
are shown in Fig.~\ref{fig: n2m}. 

Our central observation is that these plots of the structure factor 
-- from the high-temperature Monte Carlo simulations, the ideal lattice model 
without doubly-charged monopoles, and the single-tetrahedron calculation -- 
resemble each other strongly. 
We can thus use the analytical result from the latter to describe the rods,
and the physical picture provided by the lattice model to identify the single
most important ingredient for the genesis of the rods: the singly-charged
monopoles do indeed remove the pinch points; but the suppression of the 
doubly-charged ones leaves the rods behind. 
%
%
%
%

%
%

Let us now analyse the form of the correlations in more detail. The
calculation (iii), presented in App.~\ref{sec: single tet calc}, yields 
\begin{eqnarray}
{\mathcal F}_{\rm s.t.}(\vec{k}) 
&\propto& 
	\frac{1}{k^2}
	\Big[
	  k_x^2 c_y c_z
	  +
		k_y^2 c_z c_x 
	  +
		k_z^2 c_x c_y 
\nonumber \\ 
&& \quad 
    + 
		k_x k_y s_x s_y 
    +
		k_y k_z s_y s_z 
		+
		k_z k_x s_z s_x 
	\Big]
\, , 
\label{eq: NS single tet 2mono only}
\end{eqnarray}
where $c_\alpha = \cos\left(\pi k_\alpha / 2 \right)$, 
$s_\alpha = \sin\left(\pi k_\alpha / 2 \right)$, and the vectors $\vec{k}$ 
are expressed as customary in reciprocal lattice units (r.l.u.). 

\begin{figure}[h]
\includegraphics[width=0.95\columnwidth]{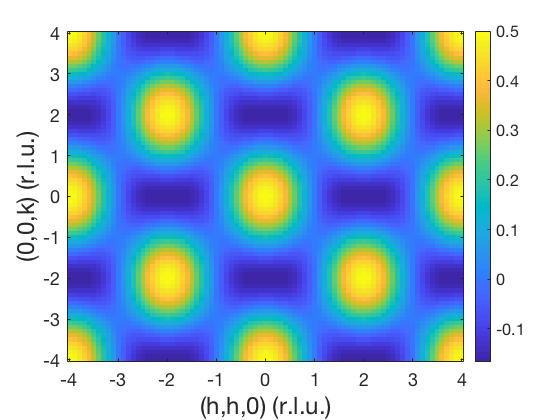}
\\ 
\includegraphics[width=0.95\columnwidth]{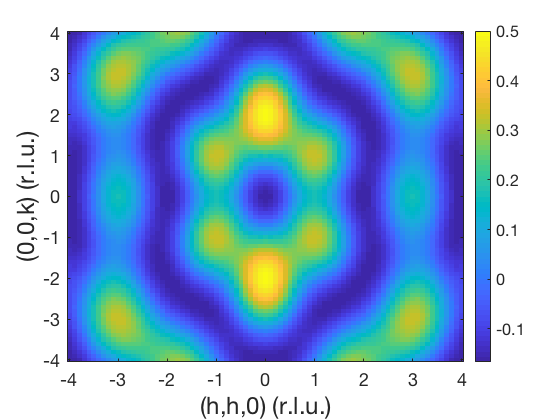}
\caption{
\label{eq: SdotS single tet 2mono only} 
Single tetrahedron calculation of the structure factor 
summed uniformly over all configurations but the all-in and all-out ones. 
Here we computed separately the two contributions to the neutron scattering 
transverse projector in Eq.~\eqref{eq: NSF full}, 
shown respectively in the top and bottom panel. 
The delta function contribution (top panel) corresponds to the conventional 
spin correlator, 
$\langle \vec{S}(\vec{q}) \cdot \vec{S}(-\vec{q}) \rangle$.  Their sum
gives the rods displayed in the right panel of the previous figure.
}
\end{figure}

In particular in the [001] direction, $\vec{k}=(0,0,k_z)$, 
Eq.~\eqref{eq: NS single tet 2mono only} simplifies to a constant, 
which in real space is of course just a delta function describing
uncorrelated planes. 
This, however, is not the full story. 
Replacing the neutron scattering transverse projector, linking spin
correlations to the neutron scattering cross section, with a delta 
function yields the 
$\langle \vec{S}(\vec{q}) \cdot \vec{S}(-\vec{q}) \rangle$ 
structure factor in the left panel of 
Fig.~\ref{eq: SdotS single tet 2mono only}. 
The rods have disappeared, and the correlations in the
[001] direction look no weaker than those in other
directions. It just so happens, for the directions of the 
rods, that the  correlations are purely longitudinal, and therefore
invisible to the neutrons on account of the transverse projector
which comes along with the scattering matrix elements. 
While the three-dimensional nature of the correlations in the presence
of scattering rods is therefore -- of course -- 
not at variance with any principle of
physics, we believe that this is a rare instance in which a scattering
rod does not go along with the emergence of $d-1$ dimensional correlations.
In a larger view of reciprocal space, this is  directly evident:
the rods do not get replicated in higher Brioullin zones. 

Next, we turn to the broader implications of the relative success of
the simple model calculation based on a single tetrahedron. A priori,
one would expect a single tetrahedron calculation to have a chance of
being accurate if correlations beyond n.n. distance are small. This
is obviously the case at high temperatures, where the leading term in
a series expansion is just 
\begin{eqnarray}
\langle S^\alpha_i S^\beta_j \rangle \propto -H^{\alpha\beta}_{ij}/T 
\, . 
\label{eq:hte_corr}
\end{eqnarray}
Hence, to the degree that the correlations remain
short-ranged, one might hope that a single-tetrahedron based
calculation also remains accurate.

There is a further simplification for spin
ice. For a single tetrahedron, the average over single-monopole states
encodes no pair correlations, as half the bonds are frustrated and
half satisfied in each state. In addition, the average over all
$2^4=16$ configurations, corresponding to infinite temperature, also
vanishes. This implies that the all-in and all-out correlations
are just the negative of those over the six ice-rule obeying
configurations. The former in turn is closely related to simply the
adjacency (i.e., the Hamiltonian) matrix of the interaction cluster. 

In other words, the presence of the monopoles effectively cuts off the
correlations at nearest-neighbour distance; from where the only form
of the correlations allowed by symmetry are those given by the
adjacency matrix -- thence the relation between correlations and the
Hamiltonian matrix in Eq.~\eqref{eq:hte_corr}. 
%
%

{\it Conclusions and outlook.} 
Our simple insight is that rod-like motifs become more pronounced as single 
monopole excitations are mixed into the 2in-2out states, while double 
monopoles remain sparse. We expect the following ingredients to be conducive 
to generating such a situation. 
(i) At a purely classical level, the energy cost of double monopoles is four 
times larger than single monopoles, and their relative Boltzmann suppression 
produces an intermediate 
temperature regime where single monopoles are relatively dense 
while double ones are negligible; this corresponds to the broadening of pinch 
points at finite temperature in classical spin ice 
compounds~\cite{Fennell2009}. 
(ii) The spin flip matrix elements in inelastic neutron scattering favour 
at low temperatures fluctuations between 2in-2out states and 3in-1out and 
3out-1in states, whereas the appearance of 4in and 4out states is in a sense 
suppressed as a higher order process. 
(iii) Finally, quantum fluctuations result in a kinetic energy gain linear
in transverse (`quantum') terms in the Hamiltonian for the 
monopoles, which is not present at this order
in tetrahedra satisfying either 
2in-2out or all-in or all-out correlations. This reduces the gap to 
single monopoles with respect to the gap to double monopoles 
(see also Ref.~\onlinecite{Kato2015}). 
The enhanced density of single monopoles at intermediate energies 
shows up in inelastic neutron scattering in quantum spin ice,
which accounts for the rod-like motifs in Fig.~\ref{fig: rods literature}. 

In real compounds, inevitably, 
departures from ideal rods will be present, as a function of 
residual longer-range correlations.
In particular, for long-range dipolar 
interactions, the size of which is appreciable especially for the large-spin 
canonical classical spin ices, 
pinch-point correlations persist~\cite{Sen1212}, albeit 
with an amplitude that vanishes with increasing temperature. 
The intermediate temperature range mentioned above will correspondingly 
exhibit some modulation along the rods,
see e.g. Refs.~\onlinecite{Fennell2009,Chang2010}. 

Beyond this, in breathing pyrochlores, the exchange coupling can differ 
significantly between the two tetrahedral sublattices ($J_1$ and 
$J_2$)~\cite{Okamoto2013,Tanaka2014,Saha2016,Lee2016}. 
Therefore, in a temperature regime where 
$J_1 \ll T \ll J_2$, the system is strongly correlated within tetrahedra 
in sublattice $2$ and largely decoupled across tetrahedra in sublattice $1$. 
There, we expect to observe well-formed rods (or the negative 
thereof) for ferromagnetic (antiferromagnetic) $J_2$ interactions. 

In summary, we have accounted in a simple fashion 
for the appearance of rods in the neutron scattering data
in a wide variety of settings in spin ice compounds and models.
This includes the identification of a notable identity between the 
structure factor obtained at high temperature and the one from an 
uncorrelated sum over single tetrahedron lowest energy configurations. 

In a broader context, it is natural to ask whether this 
also has implications for spins beyond the easy
axis limit. 
Here, preliminary results indicate that this correspondence extends to 
a broader class of frustrated magnetic models.
At the same time, the appearance of rods is not the only way for the 
pinch-points to disappear. In frustrated continuous spin systems, it has 
recently been realised that another way of  filling in  the pinch-points 
leads to a different characteristic motif, namely that of half-moon pairs 
whose radii are energy-dependent~\cite{ShannonMoon,JaubertUdagawaMoon}. 

Between them, these results not only provide signatures of the topological 
magnetism -- and its disappearance -- 
underpinning spin ice with its emergent gauge field, but they also establish 
such systems as enriching our collection of characteristic motifs -- beyond 
Bragg peaks and broad paramagnetic features -- in neutron scattering. 
%
%

\textit{Acknowledgements.} 
We are grateful to Akshat Pandey and Peter Holdsworth for useful discussions. 
This work was supported in part by EPSRC Grant No.\ EP/K028960/1 and EPSRC 
Grant No.\ EP/M007065/1 (CC) and by the 
Deutsche Forschungsgemeinschaft under grants SFB 1143 and EXC 2147 ct.qmat 
(RM). 
%
%
\appendix
%
%

\section{\label{sec: single tet calc}
Neutron Scattering details and single tetrahedron calculation
        }
We consider unit-length spins pointing along their local 
[111] easy axes, $\vec{S}_i = \sigma_i\,\hat{e}_i$, $\sigma_i = \pm 1$, 
where 
\bea
\begin{array}{l}
\hat{e}_0 = (1,1,1)/\sqrt{3} 
\\
\hat{e}_1 = (-1,-1,1)/\sqrt{3} 
\\
\hat{e}_2 = (-1,1,-1)/\sqrt{3} 
\\
\hat{e}_3 = (1,-1,-1)/\sqrt{3} 
\, . 
\end{array}
\eea
The unpolarised neutron scattering cross section (up to the single ion 
form factor, which we ignore in this work) can be written as 
\beq
\mathcal{F} = \!\!\!\!\!\!
\sum_{\alpha,\beta = x,y,z} 
\left(
  \delta_{\alpha\beta}
	- 
	\frac{k_\alpha\, k_\beta}{k^2}
\right)
\sum_{i,j} e^{\i \vec{k}\cdot\vec{r}_{ij}} 
  \langle S^\alpha_i S^\beta_j \rangle 
\, , 
\label{eq: NS full}
\nonumber
\eeq
where $\vec{r}_{ij} = \vec{r}_j - \vec{r}_i$ and $k^2 = \vert \vec{k} \vert^2$. 
The $\vec{k}=0$ point is ill-defined and should be disregarded. 
(As usual, if we were to consider polarised neutron scattering, the entirety 
of the signal would be in the spin-flip channel, whereas the non-spin-flip 
channel would give a featureless uniform background contribution.) 

It is sometimes convenient to decompose $\mathcal{F}$ into three contributions, 
a trivial one from $i=j$ ($\mathcal{F}_0$) and the two parts of the 
neutron scattering projector ($\mathcal{F}_1$ from $\delta_{\alpha\beta}$, 
and $\mathcal{F}_2$ from $-k_\alpha k_\beta / k^2$): 
\bea
\mathcal{F}_0 &=& 
\sum_{i} 
  \left[
  	1 
		- 
		\frac{(\hat{e}_i\cdot \vec{k})^2}{k^2}
	\right]
= 2 N_s / 3 
\\
\mathcal{F}_1 &=& -\frac{1}{3}
\sum_{i \neq j} 
  \cos(\vec{k}\cdot\vec{r}_{ij}) 
	\langle \sigma_i \sigma_j \rangle 
\\ 
\mathcal{F}_2 &=& -\frac{1}{k^2} 
\sum_{i \neq j} 
	(\hat{e}_i\cdot \vec{k})(\hat{e}_j\cdot \vec{k})
  \cos(\vec{k}\cdot\vec{r}_{ij}) 
	\langle \sigma_i \sigma_j \rangle 
\, , 
\eea
where $N_s$ is the total number of spins in the system. 

The Fourier transforms are taken with respect to the fcc lattice formed by 
one sublattice of tetrahedra. A conventional basis for this lattice can be 
written as 
$\vec{a}_1 = \ell (1,0,1)/2$, 
$\vec{a}_2 = \ell (1,1,0)/2$, and 
$\vec{a}_3 = \ell (0,1,1)/2$, 
where $\ell$ sets the unit length of the lattice (notice that, with this 
choice of normalisation, $\ell$ is the side of the cubic unit cell of the 
lattice). 
The reciprocal lattice basis vectors are 
$\vec{b}_1 = 2\pi (1,-1,1)/\ell$, 
$\vec{b}_2 = 2\pi (1,1,-1)/\ell$, and 
$\vec{b}_3 = 2\pi (-1,1,1)/\ell$. 
In real space, the nn distance on the pyrochlore lattice 
$r_{\rm nn} = \ell/\sqrt{8}$ is customarily set to $1$. 
However, in reciprocal space it is customary to work in so called 
reciprocal lattice units (r.l.u.), namely $\ell = 2\pi$, and we 
adopt this convention in the results presented in this work. 

In a system of $L \times L \times L$ unit cells, the available points in the 
Brilloun Zone are spanned by 
$\vec{k} = m \vec{b}_1 /L + n \vec{b}_2 /L + p \vec{b}_3 /L$, 
with $m,n,p=0,...,L-1$. 
Namely, 
\beq
\vec{k} = \frac{1}{L}
\left( 
  m+n-p, \: -m+n+p, \: m-n+p 
\right)
\, . 
\eeq
If we focus on the $(h,h,k)$ plane, then $m=p$ and we are restricted to the 
points $\vec{k} = (n, n, 2m-n)/L$. It is then convenient to forego half of the 
points on the plane in exchange for a fully symmetric grid, and choose 
$n=2i$ and $2m-n=2j$, whereby $\vec{k} = 2 (i,i,j)/L$. This is the choice 
adopted in all the Monte Carlo data presented in this work. 

For a single tetrahedron, $N_s=4$ and the $2^4=16$ configurations can be 
divided into three subsets: 
(i) 6 2in-2out states, where 
$\langle \sigma_i \sigma_j \rangle = -1/3$; 
(ii) 8 3in-1out or 3out-1in states, where 
$\langle \sigma_i \sigma_j \rangle = 0$; and 
(iii) 2 4in or 4out (aiao) states, where 
$\langle \sigma_i \sigma_j \rangle = 1$. 

The separation vectors can be written as 
$\vec{r}_{ij} = \sqrt{3} \, \pi \,(\hat{e}_j - \hat{e}_i)/4$, 
in r.l.u. units where the pyrochlore lattice constant (i.e., the tetrahedron 
side) is $r_{\rm nn} = 2\pi/\sqrt{8}$. 

Given that $\mathcal{F}=2N_s/3$ when summed trivially over all $16$ 
configurations, 
and the same is true when projected onto the 3in-1out and 3out-1in 
configurations ($\langle \sigma_i \sigma_j \rangle = 0$), then we find the 
peculiar result that projecting onto the 2in-2out ice rule states gives an 
equal and opposite structure factor than projecting onto the aiao states. 
With a few lines of algebra, one obtains the explicit expressions 
\bea
\mathcal{F}_1 &=& 
\frac{1}{6}
\left[ 
 c_x c_y + c_y c_z + c_z c_x 
\right] 
\\ 
\mathcal{F}_2 &=& 
- \frac{1}{6}
\left[ 
 c_x c_y + c_y c_z + c_z c_x 
\right] 
\label{eq: NS single tet}
\\ 
&&
+
\frac{1}{3 k^2} \left[
  k_x^2 c_y c_z 
  +k_z^2 c_x c_y 
  +k_y^2 c_z c_x 
\right.
\nonumber \\ 
&& \qquad \: 
\left.
  +k_x k_y s_x s_y 
  +k_y k_z s_y s_z 
  +k_z k_x s_z s_x 
\right] 
\, , 
\nonumber 
\eea
where $c_\alpha = \cos(\pi k_\alpha / 2)$ and 
$s_\alpha = \sin(\pi k_\alpha / 2)$, for $\alpha = x,y,z$. 
Notice that, when the two contributions are summed together, 
$\mathcal{F}_1$ cancels exactly the first term in $\mathcal{F}_2$ and 
therefore, up to a trivial overall constant, 
the single tetrahedron neutron scattering structure factor is given by the 
last two lines in Eq.~\eqref{eq: NS single tet} only. 
%
%

\end{document}